\documentstyle[aps,graphicx, times, floats]{revtex}
\graphicspath{{/home/ykao/PAPERS/GLASS/FIGURES/}}
\DeclareGraphicsExtensions{.eps,.ps}
\begin{document}
\draft

\wideabs{

  \title{The Cuprate Pseudogap:\\Competing Order Parameters or Precursor Superconductivity} 

  \author{Jelena Stajic$^1$, Andrew Iyengar$^1$, K. Levin$^1$, B. R. Boyce$^2$ and T.R. Lemberger$^2$}
 \address{$^1$ James Franck Institute and Department of Physics, University of Chicago, Chicago, Illinois
    60637}
 \address{$^2$ Dept. of Physics, Ohio State University, Columbus, Ohio, 43210}
  \date{\today} 

\maketitle
%----------------------------------------------------------------
\begin{abstract}

In this paper we compare two broad classes of theories for the pseudogap in cuprate superconductors. The comparison in made in reference to measurements of the superfluid density, $\rho_s(T,x)$, in $YBa_2CuO_{7- \delta}$ films having a wide range of stoichiometries, $\delta$, or, hole doping, $x$. The 
theoretical challenge raised by these (and previous) data is to understand
why the T-dependence of $\rho_s(T,x)$ is insensitive to the fermionic excitation gap $\Delta(T,x)$, which opens in the normal state and persists into the superconducting state, when presumably $\rho_s(T)$ is governed, at least in part, by fermionic excitations. Indeed, $\rho_s(T,x)$ seems to have a BCS-like dependence on $T_c(x)$, which, although not unexpected, is not straightforward to understand in pseudogapped superconductors where $T_c(x)$ and the excitation
gap have little in common.  Here, we contrast ``extrinsic" and ``intrinsic" theoretical approaches to
the pseudogap and argue that the former (for example,
associated with a competing order parameter) exhibits more obvious departures
from BCS-like $T$ dependences in $\rho_s(T)$ than approaches
which associate the pseudogap with the superconductivity itself. 
Examples of the latter are Fermi liquid based schemes as well as a
pair fluctuation mean field theory. Thus far, the measured behavior of the superfluid density appears to argue against an extrinsic interpretation of the pseudogap, and supports instead its intrinsic origin.

\end{abstract}
}

One of the most interesting aspects of high temperature
superconductivity in hole-doped cuprates is its development out of a non-Fermi-liquid normal state that has a fermionic (pseudo)gap.\cite{LoramPhysicaC}  
This gap, $\Delta (T)$, is present at the onset of superconductivity and persists into the superconducting state. 
Understanding this interplay of superconductivity and the pseudogap is the 
goal of the present paper. We argue here that the most natural way to
proceed is to study a property associated with the 
superconducting phase only: the superfluid density $ \rho_s(T)$.
We do so here in the context of two different classes of theories of
the cuprate pseudogap, "extrinsic" and "intrinsic" theories, in which the normal-state gap is extrinsic to superconductivity or is a precursor, respectively. These theories are compared with systematic measurements of
$\rho_s(T,x)$, in films of $YBa_2Cu_3O_{7-\delta}$ with various hole doping concentrations, $x$. The unusual aspect of the data is that they appear to follow a universal BCS-like scaling with $T_c(x)$. This behavior is unusual because the pseudogap 
is present above $T_c$ and persists below $T_c$, and therefore must be included in the analysis. It appears at the outset that any theory that includes the pseudogap physics will predict nonuniversal behavior since the pseudogap grows as $x$ decreases, while the $T_c$ decreases.

Experiments show that the pseudogap persists below $T_c$. This 
persistence is to be understood as implying that the underlying ``normal" phase of the superconductor is distinctly different from the simple Fermi liquid phase that underlies strict BCS theory. Loram and
Tallon\cite{LoramPhysicaC} have argued phenomenologically for a persistent pseudogap on the basis of thermodynamic data.  In order to obtain a physical
meaningful entropy for the extrapolated normal state, they posit that a pseudogap is present at all $ T \le T_c$ as an additional term, $\Delta_{pg}$, in the fermionic dispersion. 
This analysis forms the basis for their inferred condensation energies.  Similarly, tunneling measurements\cite{Deutscher,Krasnov2000}, particularly the earliest STM measurements from Fischer and co-workers\cite{Renner}, have led to the conclusion that the $ T \le T_c$ normal state (e.g., measured inside a vortex core) contains a pseudogap, rather similar
to that of the normal state.

One category of theory of this persistent pseudogap has the pseudogap arising from physics \textit{extrinsic} to the superconductivity, e.g., deriving from a bandstructure effect\cite{Loram,Nozieres2,Benfatto} or from a competing order parameter\cite{Laughlin,Varma1999}. This extrinsic pseudogap approach has been formulated principally at a mean field level, which can be reasonably well justified on the basis of the fact that experiments seem to indicate only a narrow critical regime\cite{Larkin} and in some (but not all) respects these pseudogapped cuprates are surprisingly similar to conventional mean field (\textit{i.e.,} BCS) predictions, as demonstrated below. 

Another category of pseudogap theory, which we refer to as \textit{intrinsic}, has the pseudogap arising from superconductivity itself. Among these are the phase fluctuation scenario\cite{Emery}, Fermi-liquid based
superconductivity along the lines of a phenomenology set forward by
Lee and Wen\cite{Lee,Millis2}, stripe-based theories\cite{Kivelson}
as well as pair fluctuation approaches\cite{Randeriareview,KosztinChen}.

We will argue that the superfluid density provides 
a suitable basis for comparing and contrasting these different theories. Here we address extrinsic approaches with the D-Density wave (DDW) theory as a prototype,
as well as two types of intrinsic schemes: Fermi liquid based superconductivity and a pair fluctuation theory.  At the mean field level\cite{KosztinChen}, the latter will be seen to be in the same spirit as the the DDW mean field approach.

Within the intrinsic school, calculations of $\rho_s$ are frequently associated
with boson-fermion models.  Among the first such studies was the work
of Stroud and co-workers\cite{Stroud} and of Carlson \textit{et al}\cite{Carlson} which pointed out that the T-linear decrease of $\rho_s(T)$ might be due at least in part to (bosonic) phase fluctuations, rather than nodal quasiparticles in a d-wave superconductor. Subsequent work\cite{Kivelson} introduced $d$-wave fermionic excitations. Ioffe and Millis\cite{Millis3}
have argued that the Fermi liquid-based phenomenology of Lee and Wen
can be interpreted in terms of coupling between phase fluctuations
and nodal quasi-particles. Orenstein and co-workers\cite{Orenstein2}
have studied the effects of (bosonic) collective phase modes on
$\rho_s$ in the context of an inhomogeneous model for superconductivity
designed, however, for the overdoped phase.

Finally, the pair-fluctuation approach to the pseudogap \cite{KosztinChen} provides yet another boson-fermion model. This scheme is closely related to Hartree-approximated Ginsburg-Landau (GL) theory\cite{JS}, where one sees how $T_c$ is suppressed by beyond-Gaussian fluctuations or non-condensed pairs, 
which give rise simultaneously to a fermionic excitation gap
in the normal state and, thus, a breakdown of Fermi liquid theory\cite{Shina}.  In this picture, non-condensed bosons above and below $T_c$, and a fermionic pseudogap, are two sides of the same coin\cite{Patton1971}. As a consequence it follows that the excitation gap and the order parameter are different above and (at least for some range of temperatures) below $T_c$, as well.
For this reason $\rho_s(T)$ is necessarily affected by these bosonic
degrees of freedom. 

There are increasingly reports of new broken symmetries as well as
arguments for quantum critical points associated with the pseudogap
phase\cite{Varma1999,Sachdev1999} which support an extrinsic approach to the pseudogap. At the same time, observations of Nernst signals\cite{Ong2} above $T_c$ which evolve continuously into the vortex contributions
below $T_c$, and the smooth evolution through $T_c$ of the
excitation gap, provide support for a scenario in which the pseudogap is \textit{intrinsic} to the superconductivity. Thus, there is a substantial need to compare the two schools directly and here we do this below $T_c$ where the predictions are most clear cut.

\subsection{Experimental Data and Constraints Imposed}

There is a strong consensus in the field that fermionic $d$-wave quasiparticles dominate thermodynamics and transport of the superconducting state, e.g., the T-linear decrease of $\rho_s (T)$ at low T is due to thermally excited BCS-like quasiparticles near the nodes in the gap, $\Delta$. This view is natural for overdoped cuprates that do not exhibit a pseudogap above $T_c$. To extend it to underdoped cuprates, in which a d-wave gap $\Delta(T)$ turns on at a temperature $T^*$ substantially above $T_c$, and is relatively constant from $T_c$ down to $T=0$, theory must account for the pseudogap in the superconducting state. It is convenient to characterize the strength of the pseudogap below $T_c$ by an 
experimental parameter:
$ \alpha = \Delta(T_c)/\Delta(0)$. 
When $\alpha \approx 0 $ the system is ``BCS-like". $\alpha$ is of order unity in the strong pseudogap regime.

\begin{figure}[!thb]
\includegraphics[angle=0,width=3.0in] {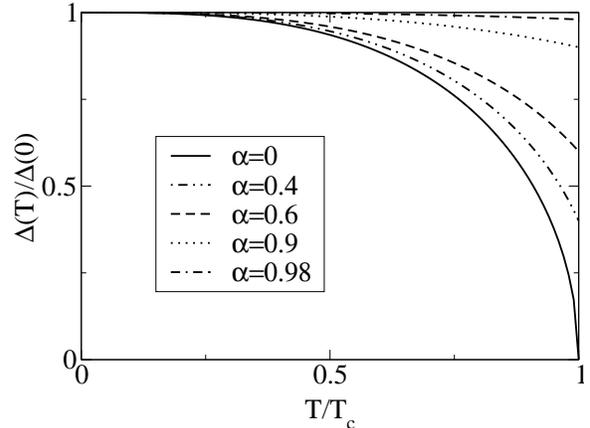}
\caption{Schematic plot of the temperature dependence of the
excitation gaps below $T_c$, for different hole concentrations.}
\label{fig:schematic}
\end{figure}

For pedagogical purposes,  in Figure \ref{fig:schematic},
we present a schematic plot of the temperature
dependence of the excitation gap within the superconducting regime, 
for a range of hole concentrations ranging from very underdoped where
$\Delta(T)$ is essentially a constant, to overdoped, where
$\Delta(T)$ follows the BCS $T$ dependence.
This plot, which can be viewed as a consolidation of an analysis, such
as that presented in Ref.\onlinecite{LoramPhysicaC}, indicates the 
role played by the important parameter $\alpha$.
This simple figure in conjunction with experimental data on
$\rho_s(T)$ to be presented below
provides a restatement of a central theme in this paper:
that, while the $T$ dependence of the
excitation gap varies dramatically as the system
becomes progressively more underdoped, nevertheless one sees rather
little change in the temperature dependence of the superfluid
density.

We make two more important points about $T_c$ and $T^*$.  
We note the experimental observation\cite{Oda,Fischer2} that the zero temperature gap $\Delta(0,x)$ is proportional to the onset temperature $T^*(x)$: 

\begin{equation}
 2 \Delta (0,x) /(k_B T^*) \approx 2.15
\end{equation}
which, because it satisfies the BCS condition, supports the idea that the pseudogap and superconductivity have a common origin. \cite{Fischer2} If so, then the challenge is to understand why $\rho_s$ vanishes so far below $T^*$. 
Another central equation which is reasonably
well satisfied by the  data presented below, is known 
as the Uemura constraint\cite{Uemura}
\begin{equation}
T_c(x) = \nu \rho_s(0,x)
\label{eq:uemura}
\end{equation}
Here $\nu$ is an unimportant
constant. Presumably this equation is intimately
connected to the physics associated with a Mott insulator.

\begin{figure}[!thb]
\includegraphics[angle=0,width=3.4in]{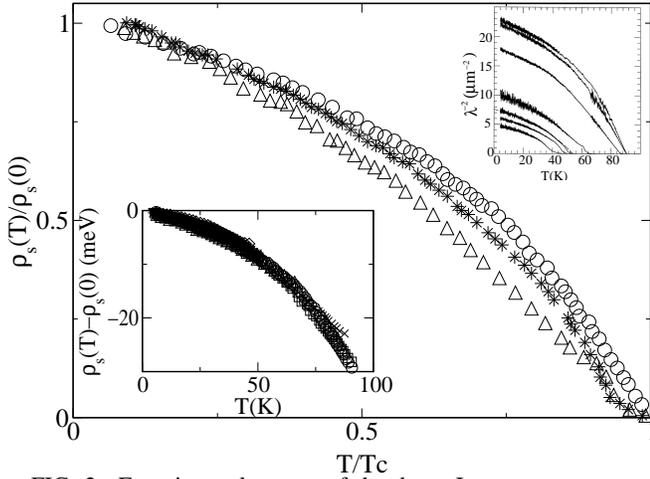}
\caption{Experimental survey of the data. Inverse square penetration depth for YBCO films with varying doping plotted in three different ways.}
\label{fig:data}
\end{figure} 

Measurements of the in-plane superfluid density in YBa$_2$Cu$_3$O$_{ 7 - \delta}$ films at various $\delta$, upper inset to Fig. \ref{fig:data}, have rather similar shapes. Films were grown by co-evaporation and consistently showed a linear low-$T$ penetration depth and inductive transition less that
$0.5$K wide as grown. Deoxygenation of the films was accomplished
by heat treating the films in an Ar atmosphere at 250$^{\circ }$C for ten
minute intervals. The sheet conductivity, $ \sigma = \sigma_1 - i \sigma_2$,
was determined from the mutual inductance of coaxial coils driven at
50 kHz located on opposite sides of the film. $\sigma_1$ is much smaller than $\sigma_2$ everywhere except close to $T_c$. From $\sigma_2$ we define $\lambda^{-2} = \mu_0 \sigma_2 \omega / d $, where $d$ is the film thickness. For purposes of later comparison with calculations, 
we define $\rho_s \propto \lambda^{-2}$ to have units of energy: $\rho_s \equiv {\hbar^2 d_c \lambda^{-2 }/ 2e^2 \mu_0}$. $d_c$ is the c-axis lattice constant in YBCO, about 1.17 nm.

The similarity of behavior for different oxygen concentrations can be seen 
either by plotting normalized data, of which three typical curves are shown in the main portion of Figure \ref{fig:data}, or by plotting the change in superfluid density, $\rho_s (T) - \rho_s (0)$ vs. $T$, lower inset. In brief, the quasiuniversal aspects of the data are that the low-T slope changes little with underdoping, and the curvature at low T remains negative and small. These results are consistent with earlier experiments\cite{Panagopoulos1} on LaSrCuO as well as with single crystal YBaCuO data at two different oxygen stoichiometries\cite{Hardy2}. Results such as these which show little difference between the overdoped (or optimal) samples where BCS theory is expected to apply, and underdoped samples, are the principal reason for the widespread belief that below $T_c$ pseudogap effects disappear, and the material becomes an ordinary BCS superconductor. The only energy scale evident in the data is $T_c$, in spite of the growing pseudogap with underdoping.

Finally, we note that data on films do not show evidence for a significant critical region near $T_c$ for any $x$. The underdoped films exhibit some inhomogeneity near $T_c$, so the experimental
case for them is not strong, but the transition in the optimally doped film is less than 0.5 K wide, and for it the case is strong. In the end, we assume that it is reasonable to examine the data in the light of theories that are not too far from mean-field.
It should also be noted that the 
universality in $\rho_s(T)$ is evident at \textit{all} temperatures.  This
is most apparent in the off-set plot, but this and the rescaled
plot suggest that the behavior around $T_c$ is as important, and
needs to be understood as systematically as the behavior around
$ T= 0$.  
On the basis of this universality, one might also conclude 
that some form of mean field theory
which does not stray too far from the BCS picture is a reasonable
starting point for addressing these $\rho_s(T)$ data.
This viewpoint provides the background for some of the theoretical
discussion which follows.

\subsection{Fermi Liquid Based Superconductivity}

\begin{figure}[!thb]
\includegraphics[angle=0,width=3.4in,height=2.6in]{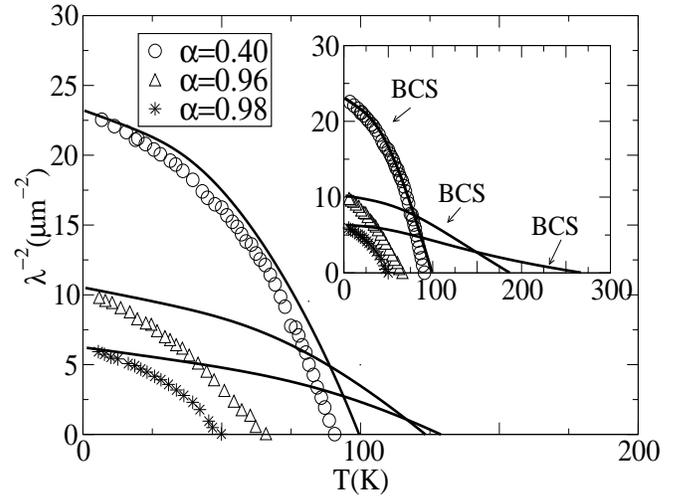}
\caption{Comparison between Fermi liquid based theories without Landau parameters and experiment for three values of $x$. 
Main figure: includes the Uemura constraint, as discussed in text.
Upper right inset: theoretical prediction for strict BCS theory.}
\label{fig:fermiliquid1}
\end{figure}

In the inset of Figure \ref{fig:fermiliquid1} we present predictions for
$\lambda^{-2}(T)$ based on BCS theory, given that $T^*$ is the
transition temperature, and given the experimental\cite{LoramPhysicaC}
behavior of the magnitude of the excitation gap. The solid lines
labelled ``BCS" plot the BCS predictions for $\lambda^{-2}(T) \propto
\rho_s ^{BCS} (\Delta(T))$, for three values of $x$, with indicated
values of the parameter $\alpha$, introduced above. Here the momentum dependence of the $d$-wave
gap was chosen to nearly fit the slope in the optimal case ($\alpha
=0.4$). The same gap shape was then used for all $x$. The bandwidth $t(x)$ was chosen to agree with the measured $\lambda^{-2}(x,0)$.

This disagreement between experiment and BCS theory, as applied in this plot, might be explained within three different scenarios. In the extrinsic pseudogap model the BCS curves are inappropriate since $T^*$ and the normal state excitation gap have nothing to do with superconductivity. 
Alternative Fermi liquid based theories 
of the superconducting state\cite{Lee,Millis2} 
belong to the intrinsic school which presumes that
the pseudogap derives from superconductivity. They introduce Landau
parameters  to fit the low $T$ slope. 
Finally, one might argue that there are other forms of excitation
of the condensate (i.e., bosonic pair fluctuations) 
which form the basis for a mean field theoretic 
intrinsic approach\cite{KosztinChen} to be
discussed later.  

In BCS d-wave theory with a tight-binding band model with nearest-neighbor hopping, $t$, the low-$T$ behavior of $\rho_s$ is given by:
\begin{equation}
[\rho_s(T)- \rho_s(0,x)]^{BCS}
\approx -\frac{4 t}{\Delta (0,x)} \frac{\ln 2}{\pi} T 
%- A \sqrt{T} 
\end{equation}
Fermi liquid based theories introduce Landau parameters $\gamma _{FL}$
which modify the low temperature slope:
\begin{equation}
[\rho_s(T)- \rho_s(0,x)]^{FL}
\approx -\frac{4 t}{\Delta (0,x)} \frac{\ln 2}{\pi} \gamma _{FL}T 
%- A \sqrt{T} 
\label{eq:rhos_FL_lowT}
\end{equation}
If it is presumed that $\rho_s(0) \propto T_c(x)$ and furthermore, that (unlike in the BCS case) $\rho_s(0)$ is independent of the hopping integral $t$,
one arrives at a form in which $T_c(x)$ now appears explicitly. 

\begin{equation}
\left[\frac{\rho_s(T)}{\rho_s(0,x)}\right]^{FL} \approx 1 - \gamma_{FL} 
\frac {t \nu } {\Delta(0,x)} \frac{4\ln 2 }{\pi} \frac {T} {T_c(x)}
\label{eq:rhos_FL_lowT_rescaled}
\end{equation}

In order for the right hand side of this equation to depend only on
$T/T_c(x)$, $\gamma_{FL}$ must counter the strong $x$ dependence of 
$\Delta(0,x)$, or equivalently, $T^*$. Much of the literature has concentrated on $x$-dependences in the $d$-wave function shape\cite{Mesot,Millis3,Taillefer} which are presumed to be responsible for some of the cancellation of $x$-dependences in the prefactor of $T/T_c(x)$ in Eq \ref {eq:rhos_FL_lowT_rescaled}. We are inclined to view this as a peripheral effect since trends in the $\rho_s$ data presented here are so strikingly systematic. Furthermore, a 
microscopic mechanism must be identified
to suppress $\rho_s$ to zero well below $T^*$. Strong thermal phase fluctuations have been suggested as a possibility.
%***I do not understand the next couple of paragraphs. I thought that Tc(x) was the measured Tc, yet calculated curves go to zero well above the measured Tc. – Tom**

\begin{figure}[!thb]
\centerline{\includegraphics[angle=0,width=3.2in,height=2.5in
]{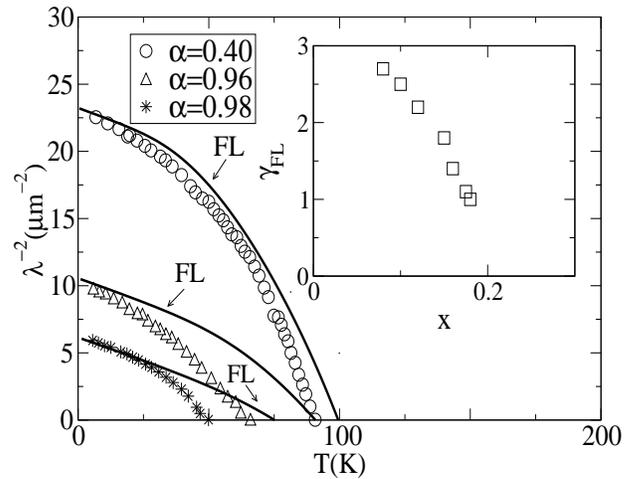}}
\caption{Fermi liquid based fits to $\rho_s$ for three values of $x$ with Landau parameters inserted to fit the low $T$ slopes. The appropriate values of 
Landau parameters for all seven samples of Figure \ref{fig:data} are shown in the inset.}
\label{fig:fermiliquid2}
\end{figure}

The main frame of Figure \ref{fig:fermiliquid1} plots this Fermi liquid result for $\rho_s$ when $\gamma_{FL} = 1$. Here to make the fitting easier, we chose slightly smaller than conventional values of $t$: $4t \approx 160 meV$ (although this is of no physical consequence) and introduced the ($x$-independent) $d$-wave shape modifications discussed in the first paragraph of this section, so that at optimal doping, agreement with experiment is reasonable. It can be seen that because of the way the Uemura relation is enforced, the discrepancies between theory and experiment are less than for the strict BCS predictions. Choosing appropriate values of $\gamma_{FL}$ as shown in the inset of Figure \ref{fig:fermiliquid2} leads to precisely fitted slopes and less discrepancy at all $T$, as plotted in the main frame of Figure \ref{fig:fermiliquid2}.
It should be noted from this Figure, however, that once Fermi liquid
parameters are introduced to fit the low $T$ slope, there are discrepancies
in the region around $T_c$. That the calculated curves go to
zero well above the measured $T_c$ is presumably a consequence of the
fact that phase fluctuations need to be invoked in this temperature
regime. 

Fermi-liquid based approaches, thus, obtain the measured universal slope for the rescaled curves by introducing\cite{Millis2} $x$-dependent Landau parameters.
These approaches also presume a strong interconnection between the universality found in both the rescaled and off-set plots of $\rho_s(T)$ and the Uemura relation. There have been a variety of experimental studies\cite{Mesot,Taillefer} which build on this Fermi liquid picture. Indeed, this approach represents the most natural extrapolation of our BCS-based intuition. The theoretical work of Ioffe and Millis\cite{Millis3} suggests that these Landau parameters arise from coupling to phase fluctuations, and, in view of the systematic $x$-dependences inferred above, it might be reasonable to speculate that coupling to these phase fluctuations becomes progressively more important with 
underdoping.

\subsection{Generalized Mean Field Theories of the Cuprate Pseudogap}

There are two other approaches to pseudogap physics
which are based on mean field theoretic approaches. In intrinsic models, $T^*$ is taken to be the mean-field superconducting
transition temperature, and 
either phase fluctuations (as in the Fermi liquid based theories,
we have discussed above)
or pair fluctuations (to be discussed below)
suppress $\rho_s$ to zero at a temperature well below $T^*$. 
In extrinsic theories $T^*$ marks the onset of another order parameter.
Some support for adopting a mean field theoretic approach comes
from the fact that $T^*$ and $T_c$ can be orders of magnitude
apart, so that it seems reasonable to establish an improved mean
field theory, and then append fluctuation effects as they
appear appropriate.  
In extrinsic models the pseudogap arises from physics other than superconductivity. Both extrinsic and intrinsic classes of
mean field theories account for $T_c \ne T^*$ from the outset, 
and the behavior of $\rho_s$ is at \textit{low} temperatures already 
intimately connected to the $x$-dependent pseudogap physics responsible for 
the separation of the higher temperatures $T_c$ and $T^*$, as will be
discussed below.

In both the intrinsic or extrinsic mean field
schools, the \textit{superconducting pseudogap state} ($T \le T_c$) is associated with the generalized equations for the gap and chemical potential $\mu$  

\begin{equation}                
g^{-1} +  \mathop{\sum_{\bf k}} \frac{1 - 2 f(E_{\bf k})}{2 E_{\bf k}}\varphi _{\bf k}^2 = 0
\label{eq:gap_equation}
\end{equation}

\begin{equation}
n=2 \sum _{\bf k} \left[ f(E_{\bf k})+v_{\bf k}^2(1-2 f(E_{\bf k}))\right]
\label{eq:number_equation}
\end{equation}

\noindent where $n$ is the electron filling factor, $\varphi_{\bf k}$
represents a general $d$-wavefunction shape, and $g$ the superconducting coupling constant. The quantity $v_{\bf k}$ is the coherence factor (described below) and $E_{\bf k}$ is the fermionic excitation energy which depends on the superconducting order parameter $\Delta_{sc}$ and the pseudogap energy scale $\Delta_{pg}$. The extrinsic pseudogap is associated with the mixing of different ${\bf k}$ states, leading to the dispersion:
\begin{equation}
E_{\bf k} ^{extrinsic}  = \sqrt{(\epsilon_{\bf k} ^ {pg} - \mu)^2 + \Delta_{sc}^2(\bf k)}
\label{eq:Extrinsic_dispersion}
\end{equation}
where\cite{Laughlin,Benfatto,Nozieres2,discussnu2}
$\epsilon_{\bf k} ^{pg}= - \sqrt {\xi_ {\bf k}^2 + \Delta_{pg}({\bf k}) ^2 }$ 
%(with $\nu = \pm 1 $), and the bare dispersion is: $\xi_{\bf k} = -2t (\cos k_xa + \cos k_ya)$. The ${\bf k }$ dependencies of $\Delta_{sc}$ and $\Delta_{pg}$ are given by $\Delta_{sc,pg}({\bf k}) = \Delta_{sc,pg} \varphi_{\bf k}$.

While there are a variety of different intrinsic theories of the normal state pseudogap\cite{Emery,Lee,KosztinChen} only the last of these\cite{KosztinChen} is readily compared with the extrinsic model discussed above, principally because it, too, represents a mean field treatment in which the two scales $\Delta_{sc}$ and $\Delta_{pg}$ appear. A detailed theoretical basis for this theory is given in a series of papers (See Ref.\onlinecite{KosztinChen} and references therein). The principal assumption is a ground state wavefunction of the BCS form\cite{Leggett} with arbitrarily strong coupling constant $g$ (which is parameterized via $T^*/T_c$ or equivalently $\alpha$) and self consistent $\mu$ so that Eqs. (\ref{eq:gap_equation}) and (\ref{eq:number_equation}) apply.

Below $T_c$ this strong coupling leads to pairing fluctuations of very low frequency and momentum contributing a fermion self-energy similar to the superconducting self-energy, leading to a more BCS-like dispersion:\cite{KosztinChen}

\begin{equation}
E_k ^{intrinsic} = \sqrt{(\xi_{\bf k} - \mu)^2 + \Delta^2({\bf k})} 
\label{eq:Intrinsic_dispersion}
\end{equation}
where $ \Delta^2 ({\bf k}) = \Delta_{pg}^2({\bf k}) + \Delta_{sc}^2({\bf k})$.
%$\xi_{\bf k}=-2 t(\cos k_x a +\cos k_y a)$ is the bare band dispersion.
We may now write the coherence factors and quasi-particle velocities
in consolidated notation; these are given by $v_{\bf k}^2 = 
\frac{1}{2}( 1 - (\epsilon_{\bf k}-\mu)/ E_{\bf k})$, and
$\partial_{\bf k} \epsilon_{\bf k}$, \cite{discussnu2} with
$\epsilon_{\bf k}= \xi_{\bf k }$ and $\epsilon_{\bf k }^{pg}$ for the intrinsic and extrinsic cases, respectively. Interestingly, a substantial body of evidence\cite{LoramPhysicaC} for a purportedly \textit{extrinsic} pseudogap comes from assuming this \textit{intrinsic} dispersion\cite{Loramwords}.

\begin{figure}[!thb]
\centerline{\includegraphics[angle=0,width=3.2in,height=4.0in
]{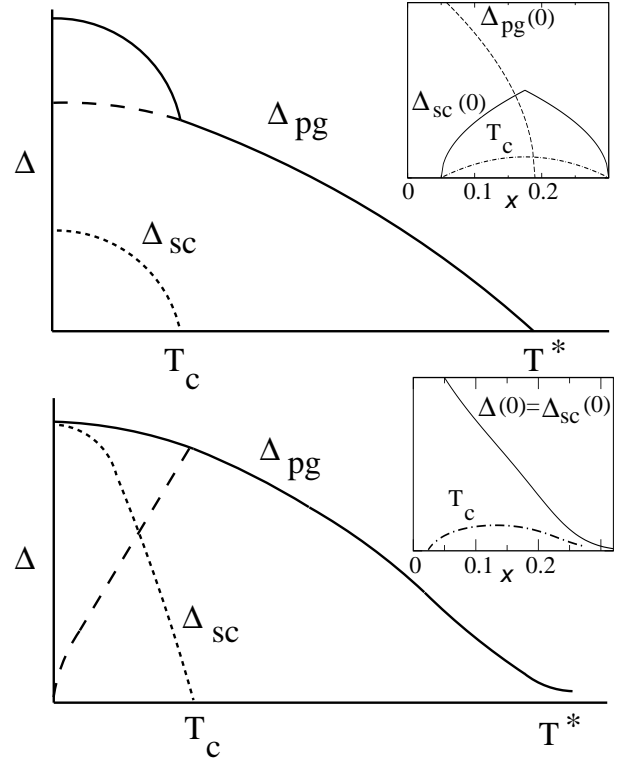}}
\caption{Schematic energy gaps for extrinsic (upper panel) and intrinsic (lower
  panel) case.  Dashed
lines show $\Delta_{pg}$ below $T_c$, dotted lines are the
superconducting order parameter while full lines represent $\sqrt {\Delta _{sc}^2+\Delta _{pg}^2}$.  Insets indicate the $x$ dependence of the gaps at $T=0$.}
\label{fig:gaps}
\end{figure}

The schematic $T$ and $x$ dependences of the various energy gaps in the two
scenarios are sketched in Fig. \ref{fig:gaps}. For the extrinsic case
(upper panel) superconductivity forms on top of a pre-existing pseudogap in the normal-state dispersion, which first appears at $T^*$ and is weakly $T$-dependent below $T_c$. For the intrinsic case (lower panel) $T^*$ marks a gradual onset of the pseudogap associated with pairing fluctuations, which are 
to be differentiated from simple phase fluctuations
around strict BCS (mean field) theory. 
Below $T_c$, these fluctuations are similar to free bosons, with the condensed fraction ($\propto \Delta_{sc}^2$) increasing at the expense of the uncondensed fraction ($\propto \Delta_{pg}^2$) until the fully condensed $T=0$ ground state is reached\cite{KosztinChen}. In effect, the normal-state pseudogap evolves smoothly into the superconducting energy. As for the $T$-dependence of the crossover, the bosonic degrees of freedom behave, to leading order, as an ideal Bose gas, so the pseudogap portion of the total gap decreases as: 

\begin{equation}
\Delta_{pg}^2 (T) \approx \Delta^2 ( T_c) ( T/T_c) ^{3/2},\,\,\,T\leq T_c
\label{eq:phenom}
\end{equation}

This description of the bosons below $T_c$ is \textit{required}\cite{ideal}
to maintain the form of the mean field defined by Eqs.(~\ref{eq:gap_equation})
and (\ref{eq:number_equation}). While Eqs. (\ref{eq:gap_equation}), (\ref{eq:number_equation}), and (\ref{eq:phenom}) are the results of a previously discussed microscopic formalism\cite{KosztinChen}, they motivate a phenomenology which requires as input only the measured values for $\Delta(0,x)$ and $T_c(x)$. The $T$ dependence of the full gap $\Delta(T,x)$ can be expressed in terms of the BCS functional form\cite{shortcut} with ``transition
temperature'' $T^*$, as is also consistent with experiment\cite{Oda}.

The insets to Figs. \ref{fig:gaps} indicate the $x$ dependencies of
the gaps at $T = 0$. In the intrinsic case, which has more BCS-like dispersion, the gap at $T=0$ is much larger than expected from the measured $T_c$. Hence, the slope of $\rho_s$ at low T would diminish with underdoping if thermally excited fermionic quasiparticles were the only important excitation at low T. 
%Here, the narrowing %of the bandwidth $t$ (needed to accomodate 
%$\rho_s(x,0)$) with decreasing $x$ increases the effective
%coupling $g/4t$ and thus $\Delta (0,x)$.  

For the extrinsic case, while the pseudogap persists to $T=0$, the gap for superconducting excitations vanishes as $T_c$ vanishes with underdoping. Thus, these theories would predict a growing slope in $\rho_s$ at low T.

\begin{figure}[!thb]
\centerline{\includegraphics[angle=0,width=3.2in,height=2.5in
]{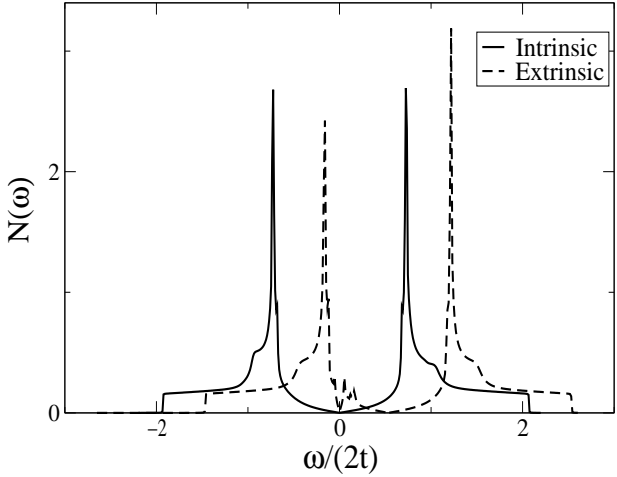}}
\caption{Comparison of densities of states for intrinsic and extrinsic schools at $T=0$ and $x=0.07$. Modified $d$-wave gap shape is applied.}
\label{fig:dos}
\end{figure} 

To further illustrate the important differences in dispersion between the two models, Fig. \ref{fig:dos} presents a comparison\cite{extr_calc}
of the density of states at $T=0$.
In the intrinsic model there are two large gap-related peaks centered on the Fermi energy. Van Hove singularities are also apparent as relatively sharp structures. In contrast, there exist two pairs of peaks in the extrinsic theory. The larger pseudogap peaks are centered around
-$\mu$, while the superconducting peaks appear around the Fermi energy.
These differences will affect transport, and in particular  $\rho_s(T)$. 

For both extrinsic and intrinsic models $\rho_s$ is of the form:\cite{discussnu2}

\begin{eqnarray}
\rho_{s_{ab}}& =&\sum _{\bf k} \frac{\Delta _{sc} ^2}{E _{\bf k}^2} 
\left[
\frac{1-2 f(E _{\bf k})}{E _{\bf k}}+2 f^\prime (E_{\bf k}) 
\right] \nonumber \\ 
& &(\partial _a \epsilon _{\bf k} \partial _b \epsilon _{\bf k} -\frac{1}{2} \partial _a \epsilon _{\bf
k}(\epsilon _{\bf k}-\mu) \partial _b\varphi _{\bf k}^2)
%+ \mbox {a small term}
\label{rho_s}
\end{eqnarray}
  
For the extrinsic case\cite{Laughlin,Nozieres2,Loram}, one finds at low $T$:
\begin{equation}
[\rho_s(T)- \rho_s(0)]^{extrinsic}
\approx -\frac{4 t}{\Delta _{sc} (0,x)} \frac{\ln 2}{\pi} T
\label{eq:rhos_extrinsic1_lowT}
\end{equation}
showing that the low-T slope should grow as $t/\Delta_{sc}(0)$, ultimately diverging at the superconductor to insulator transition. Even if we consider normalized $\rho_s$, we find
\begin{equation}
\left[\frac{\rho_s(T)}{\rho_s(0)}\right]^{extrinsic}
 \approx 1 - \frac{\nu}{\Delta _{sc}(0,x)}\frac{4 t \ln 2 }{\pi} \frac {T} {T_c(x)}
\label{eq:rhos_extrinsic2_lowT}
\end{equation}
and the same divergence.\cite{dung-hai} 
Another consequence of the extrinsic dispersion
is a $\sqrt{T}$ dependence \cite{Hae-Young}
which sets in for $T/\mu \gg \Delta_{sc}^2/\Delta_{pg}^2$.
This deviation from linearity is reflected in a convex shape for
$\rho_s(T)$. 
These results (which were derived from the DDW scheme) appear to be consequences of the fermionic dispersion which is expected to be rather general for an extrinsic pseudogap.

In the intrinsic mean-field model, the bosons or uncondensed pairs couple in a natural way to the superfluid density. Above $T_c$ (as in Hartree approximated GL theory) these pair fluctuations are responsible for the fermionic pseudogap which in turn depresses $T_c$ relative to its mean field value, $T^*$. Within the superconducting phase, these uncondensed bosons contribute to the reduction of $\rho_s$ at low $T$. Their contribution grows as the contribution of fermionic excitations decreases, as the pseudogap increases. The net result is very little apparent change in the low-$T$ behavior of $\rho_s$ with underdoping. Because of the BCS-like structure underlying the ground state (also consistent with the mean field equations (Eqs. (~\ref{eq:gap_equation}) and (\ref{eq:number_equation})), the result for $\rho_s$ is rather simple
\cite{KosztinChen,discuss_Al}
\begin{equation}
\rho_s ^{intrinsic}(T)= [ \Delta_{sc}^2  (T) / \Delta ^2 (T) ] \rho_s ^{BCS} ( \Delta (T))
\label{rho_BCS}
\end{equation}
$\rho_s^{BCS}$ is the BCS superfluid density that vanishes at $T^*$. Bosonic fluctuations simply rescale it by $\Delta_{sc}^2/\Delta^2$, which causes $\rho_s$ to vanish at $T_c$. Bosonic degrees of freedom enter this equation in that they determine the T-dependence of $\Delta_{sc}$. Rather than repeat the microscopic theory\cite{KosztinChen} presented elsewhere we introduce a more approximate but more accessible approach :
\begin{equation}
\Delta_{sc}^2 (T ) \approx  \Delta ^2 (T)[ 1 - (\Delta^2 (T_c)
 /\Delta^2 (T) ) (T/T_c) ^{3/2} ], 
\label{eq:Delta_sc}
\end{equation}
which is applicable to the entire temperature regime.
This leads to the following low $T$ dependence of $\rho_s$ 
\begin{equation}
\left[\frac{\rho_s(T)} {\rho_s(0)}\right]^{intrinsic} = 1 - [A + B(T)] \frac {T} {T_c}
\label{eq:rhos_intrinsic1_lowT}
\end{equation}
where $[A + B(T)]$ is weakly $x$ dependent and  
\begin{equation}
A = \frac {4 \ln 2}{\pi} \frac {T_c}{\Delta(0,x)}  \frac {t}{\rho_s(0)}
\end{equation}
with 
\begin{equation}
B(T) = \frac{\Delta^2 (T_c)}{\Delta^2(T)}(T/T_c)^{1/2}
\end{equation}

Here terms $A$ and $B$ correspond respectively to fermionic and bosonic excitations of the condensate. The quantity $ t / \rho_s(0)$ must be be evaluated numerically; it is found to be of order unity, and independent
of $t$. This fermionic term is precisely the same as in strict BCS theory and it reflects the full excitation gap $\Delta$.  Note that the fermionic contribution
becomes progressively smaller as the insulator is approached. On the
other hand, the bosonic contribution $B(T)$, which is only weakly $T$-dependent at low $T$, is insignificant in the overdoped regime where $\Delta (T_c)$ vanishes, and becomes progressively more important with underdoping. It can be seen that the
$x$ dependences in these two terms tend to oppose one another\cite{KosztinChen}.
In effect, this rescaled equation (Eq.(~\ref{eq:rhos_intrinsic1_lowT}))
\textit{represents a statement of the physical fact that the superfluid density is governed, even at low $T$ by its ultimate vanishing at $T_c$, as follows from the dependence on the order parameter in Eq.(\ref{rho_BCS})}. Here bosons are responsible for introducing the energy scale $T_c$ in $\rho_s(T)$.

\begin{figure}[!thb]
\centerline{\includegraphics[angle=0,width=3.2in,height=2.5in
]{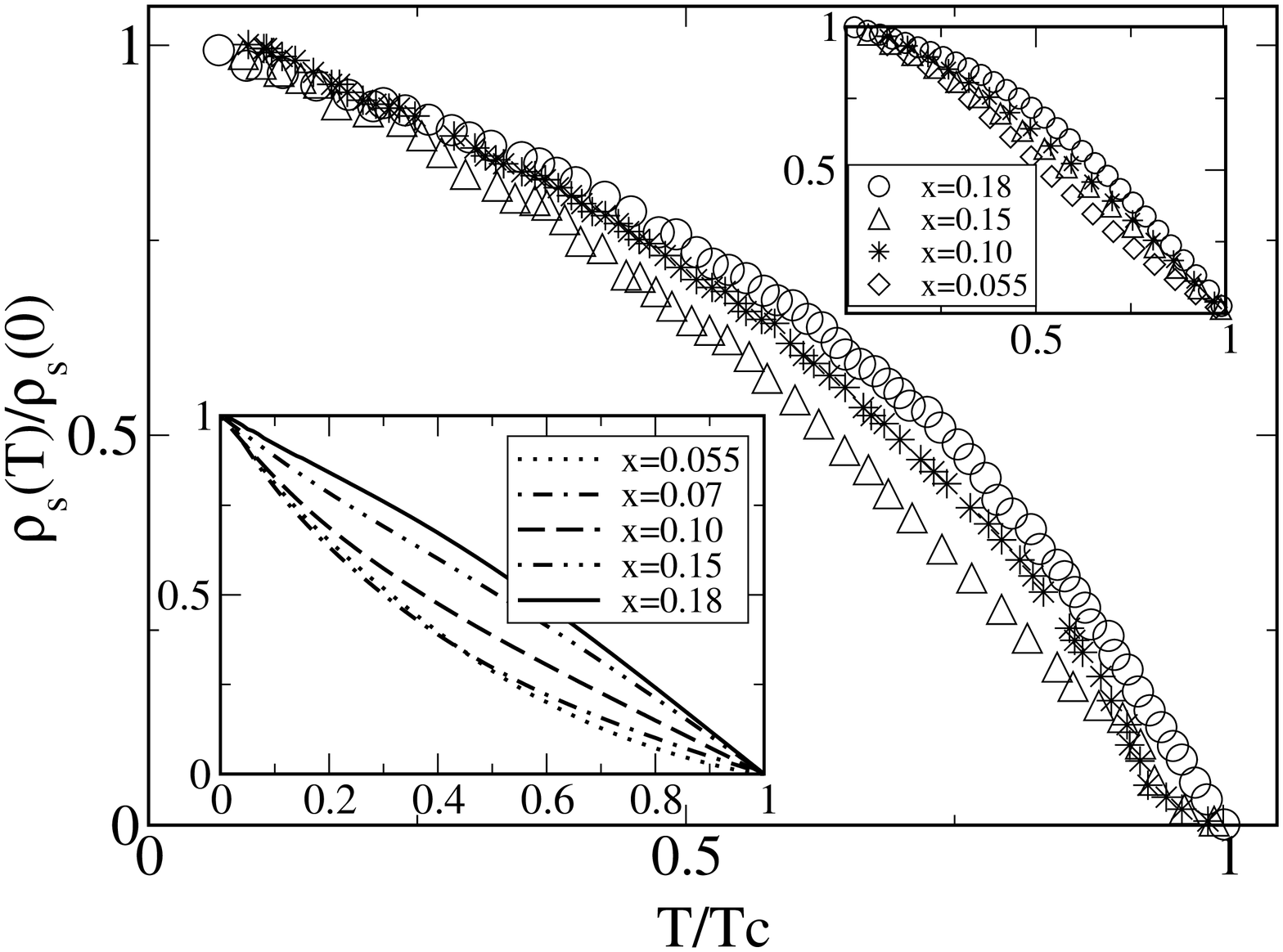}}
\caption{Rescaled plot comparing experiment (main figure) with intrinsic
(upper right) and extrinsic (lower left) mean field results.}
\label{fig:rescaled}
\end{figure} 

\begin{figure}[!thb]
\centerline{\includegraphics[angle=0,width=3.2in,height=2.5in
]{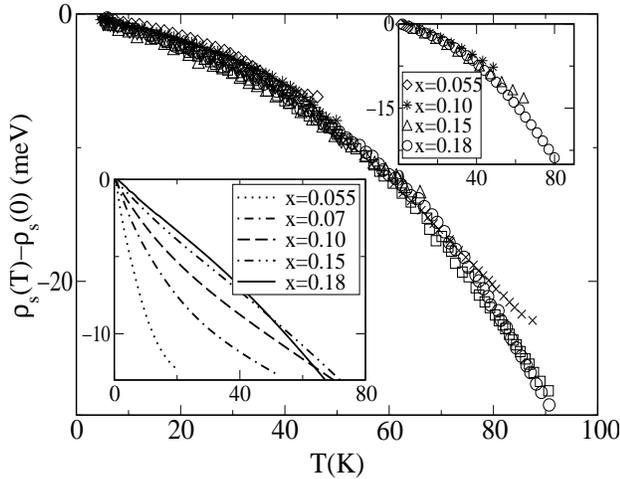}}
\caption{Offset plot comparing experiment (main figure) with
intrinsic (upper right) and extrinsic (lower left) mean field
results.}
\label{fig:offset}
\end{figure}

To examine the universality in the off-set plot of the data we may write

\begin{eqnarray}
[\rho_s(T) - \rho_s(0)]^{intrinsic}& =& 
- [A + B(T)] T~~ [\rho_s(0,x)/T_c(x)] \nonumber \\   
&\approx& -[A+B(T)]T/\nu
\label{eq:rhos_intrinsic2_lowT}
\end{eqnarray}
which shows that the off-set plots for $\rho_s(T)$ will be essentially $x$-independent only if the Uemura relation is then imposed. This will be consistent providing $t$ is taken to be $x$-dependent, which, clearly, represents an over-simplified description of Mott insulating effects. Thus, within this mean field approach, Eq.(~\ref{eq:rhos_intrinsic1_lowT}) (which does not make any assumptions about $t(x)$), is a stronger and more general result than Eq.(~\ref{eq:rhos_intrinsic2_lowT}).

We turn now to detailed numerical studies of $\rho_s(T,x)$ in both extrinsic
and intrinsic limits \cite{extr_calc}. The resulting curves for $\rho_s(x,T)$ (which can be anticipated from low $T$ expansions in Eqs. (\ref{eq:rhos_extrinsic1_lowT}), (\ref{eq:rhos_extrinsic2_lowT}),
(\ref{eq:rhos_intrinsic1_lowT}) and (\ref{eq:rhos_intrinsic2_lowT})) 
are plotted in Figures \ref{fig:rescaled} and \ref{fig:offset} for the rescaled and offset plots respectively. We can see from both Figures that for the extrinsic scenario\cite{extr_calc} the
%
%calculated slope is non-universal (diverging for low $x$)\cite{dung-hai} 
%and the 
%
general temperature dependence is progressively more convex  with underdoping, as reflects the $\sqrt T$ dependence discussed above. This, in turn, is a consequence of the position of $\Delta_{pg}$ in the extrinsic dispersion. On the other hand, the intrinsic, or strong pairing attraction theory prediction leads to concave curves, which reflect both the general concavity of the strict BCS limit and the additional $T^{3/2}$ dependence\cite{Chen3} found in Eq.(\ref{eq:phenom}). For the extrinsic case, the effects of the divergent low $T$ behavior discussed above and also noted elsewhere\cite{dung-hai} are not particularly apparent in the rescaled plot of Figure \ref{fig:rescaled}. This may reflect that fact that for the parameters chosen here we were unable to satisfy the Uemura relation.

For the off-set plots in the right hand panels the extrinsic theory
calculations illustrate the slope divergences and consequent lack of
universality, anticipated by our analytic discussions. As can be seen, the intrinsic theory leads to quite universal curves, although here (in contrast to Figure \ref{fig:rescaled}) $t(x)$ was fitted to yield the experimental $\rho_s(0,x)$. 

Because of the slope divergence and convex  curvature  at low $x$ in the extrinsic approach, at this stage the comparison between theory and experiment (in Figures \ref{fig:rescaled} and \ref{fig:offset}) favors an intrinsic origin for the pseudogap, as implemented either by the non-Fermi liquid mean field theory or by Fermi liquid based calculations with properly fitted Landau parameters. Indeed, at the level of generalized boson-fermion microscopic models for Fermi liquid     approaches\cite{Millis3}, these two schemes may ultimately turn out to contain rather similar physics.

\subsection{Conclusions}

The experiments and theories addressed in this paper are at the heart of high $T_c$ superconductivity, for they seek to unravel the nature of the excitations within the superconducting phase. To what extent are these excitations fermionic, bosonic or a mix of the two? We focus on $\rho_s(T)$, a uniquely superconducting quantity, because the experimental results are reasonably well established, \cite{Hardy2,Panagopoulos1} and the predictions of the two schools of theory differ significantly in their predictions for the evolution of the T-dependence of $\rho_s$ with underdoping. The data exhibit a quasi-universal behavior in the low-$T$ slope and curvature, that can be seen when $\rho_s(T)$ is plotted as $\rho_s(T)/\rho_s(0)$ vs $T/T_c$ or as $\rho_s(T) - \rho_s(0) $ vs $T$. Thus, it seems that $T_c$ is the only important energy scale, despite the large pseudogap in the normal state.

In the literature there are three simple ways of addressing the absence of the pseudogap energy scale in $\rho_s(T,x)$. 1) The pseudogap has an extrinsic origin, so the energy scale for fermionic excitations below $T_c$ is [see Eqs. (\ref{eq:rhos_extrinsic1_lowT}) and (\ref{eq:rhos_extrinsic2_lowT})] $\Delta_{sc}(x,T)$, which is proportional to $T_c$. 2) The pseudogap is a Fermi liquid [see Eqs. (\ref{eq:rhos_FL_lowT}) and (\ref{eq:rhos_FL_lowT_rescaled})], and Landau parameters $\gamma_{FL}$ counter the $x$-dependence in the excitation gap or equivalently\cite{Fischer2} in $T^*(x)$. Here the Uemura relationship is
essential for introducing the scale $T_c(x)$. 3) The pseudogap is associated with a strong pair fluctuations that suppress $T_c$ well below its mean field value, $T^*$. In this approach, bosonic (fluctuation) and fermionic single-particle excitations [see Eqs. (\ref{eq:rhos_intrinsic1_lowT}) and (\ref{eq:rhos_intrinsic2_lowT})] contribute to the decrease of $\rho_s(T)$ at low $T$. Here the resulting expression for $\rho_s(T)$ involves the BCS contribution with a modified pre-factor reflecting the order parameter which necessarily vanishes at $T_c$: $\rho_s(T) =[ \Delta_{sc}^2(T)/ \Delta^2 (T)] \rho_s^{BCS}(T)$.

It should be stressed that the intrinsic school pair fluctuation
approach is probably the only well established or
traditional route\cite{Patton1971} to a fermionic pseudogap
associated with homogeneous superconductivity.
Nevertheless, by
far the most widely applied intrinsic approach to the cuprate pseudogap is based on Fermi liquid theory. Much of the quantitative analysis within this scheme incorporates changes in the $d$-wavefunction shape with doping. In this
way it is inferred that the Landau parameters are considerably less $x$-dependent than would be needed to cancel the $x$ dependence in $T^*(x)$.  We have not considered this possibility in this paper since our emphasis is on qualitative issues and since the striking universality reported here for seven different $YBa_2Cu_3O_{7-\delta}$ films makes it difficult to understand how  changes in the shape of the $d$-wave gap could address this universality.

The essential contrast we have emphasized here is between two broad categories of pseudogap theories: \textit{intrinsic} and \textit{extrinsic} approaches. The
Fermi liquid approach and pair-fluctuation scheme both belong to the former.
These two may be more directly interconnected, particularly in the Fermi liquid-based interpretation of Ref. \onlinecite{Millis3}. Both start with an underlying fermion-boson Hamiltonian, where, in the Fermi liquid approach, the bosons are phase fluctuations phenomenologically coupled to the fermions, and to some extent the emphasis is more on Mott physics, and the associated softness of
phase fluctuations. By contrast the pair-fluctuation scheme builds on the short coherence length and addresses the boson-fermion coupling at a fully microscopic level. This intrinsic mean field theory is a derivative of the traditional Gaussian fluctuation picture but with two modifications: calculations are done at the Hartree level (in order to include the fermionic pseudogap) and the attractive pairing interaction is taken to be arbitrary as is, therefore, $T^*/ T_c$. Here, Mott physics enters in a more phenomenological way. Nevertheless, the bosons (which are not integrated out) are responsible for the shift in energy scale from $T^*(x)$ to $T_c(x)$. 

While we have focussed on the D-Density wave model as a prototype for the
extrinsic school, we believe it is reasonably generic and our conclusions are expected to apply to any CDW, stripe or alternative coexisting order parameter
treated at a mean field level. For this reason our results should be of
widespread interest to the community in large part because of the growing interest in quantum critical points. \textit{Because of the vanishing of $\Delta_{sc}(x)$ as the insulator is approached, extrinsic theories
are less compatible with the observed universality in $\rho_s(T)$}. Co-existing order parameters are found to lead to a breakdown of universality, and more
generally to signatures within the superconducting phase which clearly
distinguish under- and over-doped cuprates. These should be most visible near the superconductor insulator boundary. If future data on even more underdoped samples do not show significant deviations from the (thus far) ``universal" curves for $\rho_s(x,T)$,  we argue that this will provide substantial support for an intrinsic origin to the pseudogap in the context either of Fermi liquid-based or non-Fermi liquid based approaches to the superconductivity.
Making a further distinction between these two should be possible through theoretical and experimental studies of the ac \cite{Drew}, as well as thermal
conductivity, which studies are currently underway.

The discussion in this paper has focused mostly on mean field approaches.
Other, non mean field approaches can be contemplated as well, but the observation of quasi-universality in $\rho_s(T,x)$ suggests that theories of the superconducting state should not deviate too strongly from simple BCS theory \textit {at all $T$}. Additional support for this observation comes from the fact that there is often a very considerable separation (by orders of magnitude) between $T^*$ and $T_c$ which suggests that a carefully chosen mean field theory\cite{Larkin} is a more appropriate starting point for understanding the pseudogap phase than is an approach based on strict BCS physics with added fluctuation effects. One may view the off-set plot of the data in Figure \ref{fig:data} (lower left inset) as, one of the most interesting contributions of this paper and perhaps, the strongest constraint imposed by experiment on the nature of the excitations in the superconducting state. \textit{This off-set plot suggests that the behavior of $\rho_s(T,x)$ at all $T$ is a universal function and that there are no systematic differences between under- and optimally doped samples}. This, of course, is very surprising because the magnitude and temperature dependence of the excitation gap vary markedly for this range of $x$. Resolving this ``paradox" has been a central issue of this paper.

This work was supported by NSF-MRSEC Grant No. DMR-9808595 and by NSF-DMR 0203739. We acknowledge very helpful conversations with Q. Chen, Y.-J. Kao,
Q. H. Wang and H. Y. Kee.

%\bibliographystyle{prsty}
%\bibliography{/home/jstajic/References3}

%\bibliography{/home/levin/PAPERS/JELENA/References3}
%\bibliography{/home/apiyenga/JSTAJIC/References6}

\begin{thebibliography}{10}

\bibitem{LoramPhysicaC}
J.~L. Tallon and J.~W. Loram, Physica C {\bf 349},  53  (2001).

\bibitem{Deutscher}
G. Deutscher, Nature {\bf 397},  410  (1999).

\bibitem{Krasnov2000}
V.~M. Krasnov {\it et~al.}, Phys. Rev. Lett. {\bf 84},  5860  (2000).

\bibitem{Renner}
C. Renner {\it et~al.}, Phys. Rev. Lett. {\bf 80},  3606  (1998).

\bibitem{Loram}
J.~W. Loram {\it et~al.}, Jour. of Superconductivity {\bf 7},  243  (1994).

\bibitem{Nozieres2}
P. Nozieres and F. Pistolesi, Eur. Phys. J. B {\bf 10},  649  (1999).

\bibitem{Benfatto}
L. Benfatto, S. Caprara, and C. Di~Castro, European Physics Journal B {\bf 17},
   95  (2000).

\bibitem{Laughlin}
S. Chakravarty, R.~B. Laughlin, D.~K. Morr, and C. Nayak, Phys. Rev. B {\bf
  63},  094503  (2001).

\bibitem{Varma1999}
C. Varma, Physical Review Letters {\bf 83},  3538  (1999).

\bibitem{Larkin}
V. Geshkenbein, L. Ioffe, and A. Larkin, Physical Review B {\bf 55},  3173
  (1997).

\bibitem{Emery}
V.~J. Emery and S.~A. Kivelson, Nature {\bf 374},  434  (1995).

\bibitem{Lee}
P.~A. Lee and X.-G. Wen, Phys. Rev. Lett. {\bf 78},  4111  (1997).

\bibitem{Millis2}
A.~J. Millis, S. Girvin, L. Ioffe, and A.~I. Larkin, J. Phys. Chem. Solids.
  {\bf 59},  1742  (1998).

\bibitem{Kivelson}
M. Granath {\it et~al.}, Phys. Rev. Lett. {\bf 87},  167011  (2001).

\bibitem{Randeriareview}
M. Randeria,  in {\em Bose Einstein Condensation}, edited by A. Griffin, D.
  Snoke, and S. Stringari (Cambridge Univ. Press, Cambridge, 1995), pp.\
  355--92.

\bibitem{KosztinChen}
I. Kosztin, Q. Chen, Y.-J. Kao, and K. Levin, Phys. Rev. B {\bf 61},
  11662(2000); Q. Chen, I. Kosztin, B. Janko, and K. Levin, Phys. Rev. Lett.
  {\bf 81}, 4708(1998).

\bibitem{Stroud}
E. Roddick and D. Stroud, Phys. Rev. Lett. {\bf 74},  1430  (1995).

\bibitem{Carlson}
E. Carlson, S. Kivelson, V. Emery, and E. Manousakis, Phys. Rev. Lett. {\bf
  83},  612  (1999).

\bibitem{Millis3}
L. Ioffe and A. Millis, J. Phys. Chem. Solids {\bf 63},  2259  (2002).

\bibitem{Orenstein2}
J. Corson {\it et~al.}, Phys. Rev. Lett. {\bf 85},  2569  (2000).

\bibitem{JS}
J. Stajic and K. Levin (unpublished).

\bibitem{Shina}
S. Tan and K. Levin, preprint, cond-mat/0302248 (unpublished).

\bibitem{Patton1971}
B.~R. Patton, Phys. Rev. Lett. {\bf 27},  1273  (1971).

\bibitem{Sachdev1999}
M. Vojta, Y. Zhang, and S. Sachdev, Physical Review Letters {\bf 85},  4940
  (2000).

\bibitem{Ong2}
Y. Wang {\it et~al.}, Physical Review B {\bf 64},  224519  (2001).

\bibitem{Oda}
M. Oda {\it et~al.}, Physica C {\bf 281},  135  (1997).

\bibitem{Fischer2}
M. Kugler {\it et~al.}, Phys. Rev. Lett. {\bf 86},  4911  (2001).

\bibitem{Uemura}
Y.~J. Uemura, Physica C {\bf 282-287},  194  (1997).

\bibitem{Panagopoulos1}
C. Panagopoulos {\it et~al.}, Phys. Rev. B {\bf 60},  14 617  (1999).

\bibitem{Hardy2}
D. Bonn {\it et~al.}, J. Phys. Chem. Solids {\bf 56},  1941  (1995).

\bibitem{Mesot}
J. Mesot {\it et~al.}, Phys. Rev. Lett. {\bf 83},  840  (1999).

\bibitem{Taillefer}
M. Sutherland {\it et~al.}, preprint, cond-mat/0301105 (unpublished).

\bibitem{discussnu2}
If the order parameter breaks translational symmetry two bands are formed; we
  presume that $\mu$ is negative and $T$ is low, so the contribution of the
  upper- and intraband terms (and their associated velocities) are small.

\bibitem{Leggett}
A.~J. Leggett,  in {\em Modern Trends in the Theory of Condensed Matter}
  (Springer-Verlag, Berlin, 1980), pp.\ 13--27.

\bibitem{Loramwords}
It should be pointed out that an additional assumption was also made in Ref.
  ~\onlinecite{Loram}: that $\Delta(T_c)$ and $T^*$ are directly proportional.
  In this way, it was inferred that $T^*$ vanishes at a critical hole
  concentration. This is not a necessary consequence and in the (alternative)
  intrinsic scenario when $\Delta(T_c)$ is zero, one finds $T^* = T_c$.
  However, if for some reason $\mu$ is pinned to zero in these extrinsic models
  the differences in $E_k$ would be less and the behavior more BCS-like. See
  also Ref.~\onlinecite{Nozieres2}.

\bibitem{ideal}
The $T^{3/2}$ dependence of an ideal Bose gas enters because the pair
  susceptibility $\chi$ associated with Eq. (1), [which equation can be written
  as $1 + g \chi = 0$] necessarily has a $q^2$ dispersion, corresponding to a
  small $q$ expansion of $G(k-q)G_o(k)$. For details see Ref.
  \cite{KosztinChen}.

\bibitem{shortcut}
A simple parameterization of the $d$-wave $T$ dependence is $\Delta (T) \approx
  \Delta (0)\sqrt{ 1- (T/T_c)^3 }$.

\bibitem{extr_calc}
In the intrinsic case, the parameters were determined using the simple
  phenomenological approach, so that $T_c(x)$ and $\Delta(x,0)$ were fit to
  experiment\cite{LoramPhysicaC}. For the extrinsic case gap and number
  equations from Ref. \cite{Carbotte} were solved with $V_{DSC}=0.8$ and
  $V_{DDW}=2.0$. The modified $d$-wave gap shape (as in Figures 2 and 3) was
  used in both cases.

\bibitem{dung-hai}
Q.~H. Wang, J.~H. Han, and D.~H. Lee, Phys. Rev. Lett. {\bf 87},  077004
  (2001).

\bibitem{Hae-Young}
S. Tewari, H.~Y. Kee, C. Nayak, and S. Chakravarty, Phys. Rev. B {\bf 64},
  224516  (2001).

\bibitem{discuss_Al}
The difference (proportional to $\Delta_{pg}^2$) between this result and
  $\rho_s^{BCS}(\Delta)$ is due to a contribution to the response (of the
  Aslamazov-Larkin type) reflecting the direct coupling of radiation to
  uncondensed pairs of charge $2e$.

\bibitem{Chen3}
Q. Chen, I. Kosztin, and K. Levin, Phys. Rev. Lett. {\bf 85},  2801  (2000).

\bibitem{Drew}
A. Iyengar, J. Stajic, Y.-J. Kao, and K. Levin, preprint, cond-mat/0208203
  (unpublished).

\bibitem{Carbotte}
J.-X. Zhu, W. Kim, C. Ting, and J. Carbotte, Phys. Rev. Lett. {\bf 87},  197001
   (2001).

\end{thebibliography}
%\begin{thebibliography}{99}
%\bibitem{dung-hai} Q. Wang et al.: PRL {\bf 87} 077004
%\end{thebibliography}

\end{document}